\documentclass[prd,preprint,nofootinbib]{revtex4-1}
\usepackage{graphicx}
\usepackage{amsmath}
\usepackage{hyperref}
\usepackage{comment}
\usepackage{caption}
\graphicspath{{./images_resubmit/}}
\begin{document}
\textit{(Accepted for Publication in Physical Review D on January 20, 2016)}
\\
\url{http://journals.aps.org/prd/accepted/0b072Q25S231e31378663d67cfc508a50ee71ac77}

\title{Constraints on Axions and Axionlike Particles from \emph{Fermi} Large Area Telescope Observations of Neutron Stars}
\author{B.~Berenji}
\email{bberenj@calstatela.edu}
\affiliation{California State University, Los Angeles, Department of Physics and Astronomy, 5151 State University Drive, Los Angeles, CA 90032-8206, USA}
\author{J.~Gaskins}
\email{jgaskins@uva.nl}
\affiliation{GRAPPA, University of Amsterdam, Science Park 904, 1098XH Amsterdam, Netherlands}
\author{M.~Meyer}
\email{manuel.meyer@fysik.su.se}
\affiliation{Department of Physics, Stockholm University, AlbaNova, SE-106 91 Stockholm, Sweden}
\affiliation{The Oskar Klein Centre for Cosmoparticle Physics, AlbaNova, SE-106 91 Stockholm, Sweden}

\begin{abstract}
 We present constraints on the nature of axions and axion--like particles (ALPs) by analyzing gamma--ray data from neutron stars using the Fermi Large Area Telescope. In addition to axions solving the strong CP problem of particle physics, axions and ALPs are also possible dark matter candidates. We investigate axions and ALPs produced by nucleon--nucleon bremsstrahlung within neutron stars. We derive a phenomenological model for the gamma--ray spectrum arising from subsequent axion decays. By analyzing 5 years of gamma-ray data (between 60 MeV and 200 MeV) for a sample of 4 nearby neutron stars, we do not find evidence for an axion or ALP signal, thus we obtain a combined 95\% confidence level upper limit on the axion mass of 7.9$\times 10^{-2}$ eV, which corresponds to a lower limit for the Peccei-Quinn scale $f_a$ of 7.6$\times 10^7$ GeV. Our constraints are more stringent than previous results probing the same physical process, and are competitive with results probing axions and ALPs by different mechanisms.
\end{abstract}
\maketitle

\section{Introduction}

The axion is a well-motivated particle of theoretical physics.  This light pseudoscalar boson arises as the pseudo Nambu-Goldstone boson of the spontaneously broken $U(1)$ Peccei--Quinn symmetry of quantum chromodynamics (QCD), which explains the absence of the neutron electric dipole moment~\cite{axionBook,chengLi}, and thereby solves the strong $CP$ problem of particle physics~\cite{PQ,wein,wilczek}.   In addition, it is a possible candidate for cold dark matter~\cite{preskill1983cosmology, abbott1983cosmological, dine1983not}.  Astrophysical searches for axions generally involve constraints from cosmology or stellar evolution~\cite{raffelt,raffeltStellar}.  Many astrophysical studies placing limits on the axion mass have also considered axion production via photon-to-axion conversion from astrophysical and cosmological sources such as type Ia supernovae and extra-galactic background light~\cite{horns2012hardening,sanchez2009hints,brockway1996sn,csaki2002dimming}.  However, we examine a different mechanism here.  
We set bounds on the axion mass $m_a$ by considering radiative decays of axions produced by nucleon-nucleon bremsstrahlung in neutron stars~\cite{raffelt}.  %
The expected gamma--ray signal arising from this process should lie roughly between 1 MeV to 150 MeV, as a direct consequence of the axion energies produced, as  will be shown in this work.

Prior work on axions produced via nucleon-nucleon bremsstrahlung has yielded constraints on $m_a$ using X-ray emission from pulsars~\cite{axionpulsar}, and gamma-ray emission from the SN1987A remnant~\cite{giannotti}.  Here, for the first time, we use \emph{Fermi} LAT observations of neutron stars to search for signatures of axions. The \emph{Fermi} LAT detects gamma rays with energies from 20 MeV to over 300 GeV~\cite{p7rep}, and includes the range where photons from axions produced in neutron stars can be measured.   One of the advantages of our approach over previous work includes selecting multiple sources, which we combine in a joint likelihood analysis.

The neutron star sources selected for this analysis have not been detected as gamma--ray sources~\cite{2ndPsrCat},  although they have been detected in radio and X--rays as pulsars~\cite{Becker,becker2005multiwavelength,pavlov2009detection}. Pulsed emission in gamma rays would be a background to the axion--decay signal in the energy range that we consider.    Since we do not model this background, the derived limits can be regarded as conservative.

We begin with a theoretical model for axion emissivity, and derive the spectrum of axion kinetic energies numerically from the phase-space integrals for the nucleon-nucleon bremsstrahlung process.  We consider the competing process of axion conversion via the Primakoff effect (axion to photon conversion in a magnetic field).  No signal is detected, therefore we set limits on the axion mass $m_a$ by comparing the theoretical spectrum of gamma rays to the experimental constraints we obtain from \emph{Fermi} LAT observations of the selected neutron stars.    For axions, we consider the standard relation between $m_a$ and the Peccei--Quinn scale $f_a$~\cite{axionBook}:
\begin{equation} m_a \approx 6 \ \mu{\rm eV} \left(\frac{f_a}{10^{12} \ {\rm GeV}}\right)^{-1}. \label{eq:mafa} \end{equation}

We generalize our constraints to include axion--like particles (ALPs), which are light pseudo--scalar spin 0 bosons, having some axion properties.  These arise in supersymmetry, Kaluza--Klein theories, and superstring theories~\cite{witten1984some,cicoli2012type,ringwald2014searching}.  A fundamental difference between axions and ALPs is that the constraint between $m_a$ and $f_a$ in equation \eqref{eq:mafa} is relaxed, so that they are each independent parameters.

The organization of the paper is as follows.  In Section~\ref{sec:Theory}, we discuss the theory, the 	phenomenology of axion production, as well as the astrophysical model for converting the axion flux into photon flux from decays.  In Section~\ref{sec:Observations}, we present the \emph{Fermi} LAT analysis and observations of a sample of neutron stars.  In Section~\ref{sec:sim}, we discuss the estimation of the systematic uncertainties. In Section~\ref{sec:Discussion}, we discuss the implications of the results and draw comparisons with other astrophysical limits on the axion mass.

\section{Theory\label{sec:Theory}}
\subsection{Phenomenology}\label{sec:Pheno}

\begin{figure}[width=4in]
\begin{centering}
\includegraphics[width=4in]{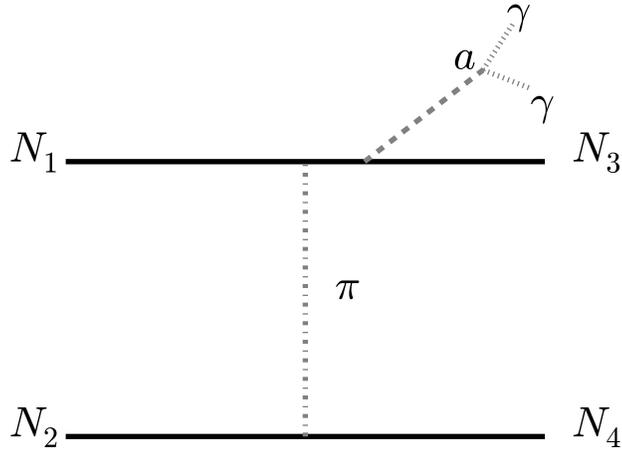}
\caption{A Feynman diagram for the nucleon--nucleon bremsstrahlung process $NN\to NNa$, according to the one--pion exchange (OPE) assumption. $N_1$ and $N_2$ are incoming nucleons, and $N_3$ and $N_4$ are outgoing nucleons.  $a$ represents the axion. $\pi$ represents a pion. In the case $nn\to nna$, we consider $\pi^0$.  We also represent the decay process $a\to\gamma\gamma$ in this diagram.}
\label{fig:Feynman}
\end{centering}
\end{figure}

Axions may be produced in neutron stars by the reaction $NN\to NNa$, where $N$ is a nucleon.  For calculation clarity, we often assume the nucleon is a neutron.   The axions produced in this manner would be relativistic (see below).  For a physical description of this process, we follow the phenomenology of Hanhart, Philips, and Reddy \cite{uw}, also described by Raffelt~\cite{raffelt}, who model the process as a nucleon--nucleon scattering process or nucleon--nucleon bremsstrahlung.  This model relies upon the well--known phenomenology of nucleon-nucleon bremsstrahlung, in the one--pion exchange approximation (OPE), which generates axions (as well as neutrinos); a Feynman diagram for this process is illustrated in Fig.~\ref{fig:Feynman}.

The axion emissivity, i.e., energy loss rate per volume, is given in natural units ($\hbar=c=1$), as~\cite{uw}:

\begin{equation} \epsilon_a = \frac{g_{\rm ann}^2}{48\pi^2M_N^2} \int d\omega \ \omega^4 S_\sigma(\omega), \label{eq:emissAxion} \end{equation}

\noindent where $\omega$ is the axion energy.
As for constants, $M_N=939$ MeV is the isospin--averaged nucleon mass, the axion--nucleon coupling is $g_{\rm ann}=C_N M_N/f_a = 10^{-8}(m_a/1 {\rm eV})$, for $C_N\simeq 0.1$. $C_N$ parametrizes the contributions from the vacuum expectation values (VEVs) of the Higgs $u$ and $d$ doublets in the axion model considered, the DFSZ model~\cite{dfsz1,dfsz2}. The DFSZ model should be distinguished from the KSVZ model~\cite{ksvz1,ksvz2}.  In the KSVZ model, the axion couples to photons and hadrons, but in the DFSZ model, axion coupling to electrons is also allowed~\cite{Iwamoto}.  

The spin structure function $S_\sigma(\omega)$ accounts for the energy and momentum transfer and includes the spins of the nucleons.  In the nucleon--nucleon scattering process, the following phase--space integral corresponding to the Feynman diagram of Figure~\ref{fig:Feynman}~\cite{uw} is defined as:

\begin{multline}  S_\sigma(\omega;\mu,T) = 1/4 \int \left[ \prod_{i=1\cdots4} \frac{d^3p_i}{(2\pi)^3} \right] (2\pi^4) \delta^3(\mathbf{p_1+p_2-p_3-p_4}) \\ \times \delta(E_1 + E_2 - E_3 - E_4 - \omega) \mathcal{F} \mathcal{H}_{ii} .
\end{multline}\label{eq:spinStruct}
In the previous equation, ${\bf p_{1,2}}$ are the momenta of the incoming nucleons,  ${\bf p_{3,4}}$ are the momenta of the outgoing nucleons, $E_{1,2,3,4}$ are the respective energies, and $\omega$ is the energy of radiated axions.  The two $\delta$--functions ensure conservation of momentum and energy.  The integration limits of the momentum variables are $0<p_i<2p_{F,n}$, where $p_{F,n}$ is the neutron Fermi momentum~\cite{Iwamoto}.  We assume non-relativistic nucleons, and take $E_i = p_i^2/2M_N$; this is justified given the neutron star temperature we assume (see below).  It can be shown that the axions are relativistic according to $p_i^2/2M_N$,  since they are produced with large Lorentz boost, even for a putative axion mass of 1 keV.\footnote{1 keV was chosen as a conservative value of the axion mass for axions with energy $\sim$ 100 MeV.} 
Equation~\eqref{eq:emissAxion} is averaged over nucleons, and the dependence on nuclear density enters through the neutron Fermi momentum via the Fermi energy (which has $n^{2/3}$ dependence on density).  In the neutron star, we assume a number density of nucleons of 0.033 fm$^{-3}$ or a mass density of 5.6$\times$ 10$^{13}$ g/cm$^{3}$.  This is within the range assumed by the perturbative approximation, where $\rho<1\times 10^{14}$ g/cm$^{3}$~\cite{raffelt}.  
We note that the  hadronic tensor function (which accounts for the nucleon spins) is approximated by $\mathcal{H}_{ii} = k/\omega^2$~\cite{uw}, where $k\sim10$ is a constant. The function $\mathcal{F}(E_1,E_2,E_3,E_4;\mu;T)$ is given by a product of thermodynamic functions:

\begin{equation} \mathcal{F} = f(E_1) f(E_2) (1-f(E_3)) (1-f(E_4)), \end{equation}

\noindent where $f(E) = 1/(1+\exp\left((E-\mu)/T\right))$.
Thus we see that $\mu$, the neutron star degeneracy~\cite{raffelt}, and $T$, the core temperature of the neutron star, are additional parameters of the model, which may vary with the neutron star source.  

We assume values of $\mu/T\simeq10$ and $T=20$ MeV.  These values are supported by the equations of state (EOS) simulations of nuclear matter of the models described in Refs.~\cite{ruster2005phase, shen1998relativistic,ravenhall}.  The neutron star temperature we use here follows the cited models, which assume relativistic conditions and beta equilibrium (a condition on the chemical potentials of neutrons, protons, and electrons) in neutron star matter.\footnote{We assume relativistic conditions in the sense of describing the interactions between nucleons.}   Since neutron stars in such models are expected to be in a superconducting phase, cooling is likely to be slower~\cite{hybridCooling} than in less-sophisticated models of neutron stars without superfluidity.  Models with and without superfluidity are compared in Ref.~\cite{Tsuruta}. This slower cooling is due to internal heating from friction between the superfluid and the neutron star crust~\cite{larson1999superfluid}, which has been investigated for J0953+0755, one of the neutron stars we examine.  In addition, observational constraints of neutron star J0953+0755 place the surface temperature at 6 eV~\cite{pavlovObs}, which may be consistent with the interior temperatures we assume.  The temperature we choose for the analysis of $T=20$ MeV is roughly the midpoint of the range of neutron star temperatures in the phase diagram for neutron stars given in Ref.~\cite{ruster2005phase}.

\begin{figure}
\begin{center}
\includegraphics[width=6in]{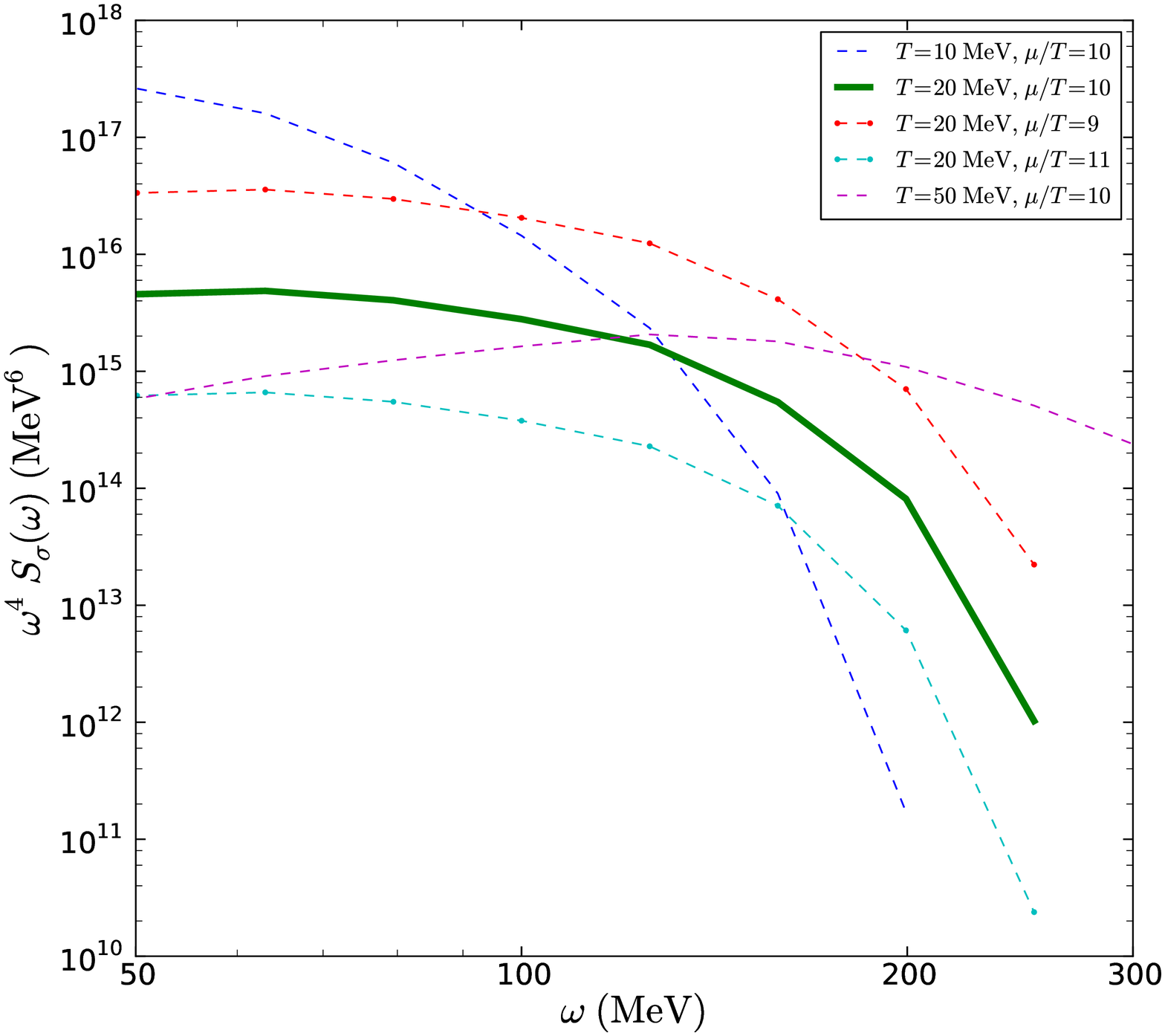}
\caption{The function $\omega^4 S_\sigma(\omega)$, which shows the energy dependence of the emissivity.  The spin structure function, $S_\sigma(\omega)$, is computed according to equation~\eqref{eq:spinStruct}, assuming non--relativistic nucleons and nucleon energy  $E = p^2/2M_N$.  We plot values of $S_\sigma(\omega;\mu,T)$ for different values of $\mu$ and $T$.  In the data analysis, we use $T = 20$ MeV, $\mu/T = 10$, which corrresponds to the solid green curve. \label{fig:spinStruct}}
\end{center}
\end{figure}

We evaluate the phase space integrals, accounting for the $\delta$--functions in energy and momentum, by numerical integration, after the analytic simplifications of Ref.~\cite{raffeltBrems}.  These simplifications are described in Appendix~\ref{appA}.   The spin structure function is plotted in Figure \ref{fig:spinStruct} for different values of $T$ and $\mu/T$.  It may be observed that increasing $T$ shifts the function to higher energies, and changes the shape of the curve, but increasing $\mu/T$ decreases the amplitude of the function for fixed $T$.

\subsection{Astrophysical Model}

We need to include factors and physical constants to convert the axion emissivity in equation \eqref{eq:emissAxion} into a gamma-ray flux (measurable with the \emph{Fermi} LAT).  In deriving a photon flux ($\Phi$), we consider the differential emissivity with respect to axion energy.  In the case of radiative decay of axions $a\to 2\gamma$, we assume for the sake of the calculation that the photon energy is simply half of the axion energy (i.e., the axion mass is negligible compared to its kinetic energy); this is justified, since in the scenario considered here, the axion is highly relativistic with respect to the observer.  In addition, we consider a neutron star of volume $V_{NS}$ as a uniform density sphere with a radius of 10 km, a timescale for axion emission $\Delta t$ to be described below, a neutron star at a distance $d$, and the axion decay width $\Gamma_{a\gamma\gamma}$ (inverse lifetime).   We consider $\Gamma_{a\gamma\gamma}$ as given by~\cite{axionBook}:
\begin{align} \Gamma_{a\gamma\gamma} = \frac{g_{a\gamma}^2 m_a^3}{64\pi} = 1.1\times 10^{-24}{\rm s}^{-1} (m_a/1 \ {\rm eV})^5 \end{align}
where $g_{a\gamma} = C\alpha/(2\pi f_a)$ is the axion--photon coupling, $C$ is an axion model parameter, and $\alpha\simeq 1/137$ is the fine structure constant.\footnote{The axion model parameter $C$ is given by $C = (\frac{E}{N}-\frac{2}{3}\frac{4+z}{1+z})$, where $E/N=8/3$ for DFSZ axions.  $z=m_u/m_d$ is the ratio of masses of the up quark to the down quark.}   

We may proceed to derive the spectral energy distribution by converting the number of axions $N_a$ emitted per unit time and unit axion energy $\omega$,

\begin{equation} \omega \frac{dN_a}{dtd\omega} = \frac{d\epsilon_a}{d\omega}  V_{NS}, \end{equation}
to the number of photons $N_\gamma$ emitted per unit time and photon energy as
\begin{equation} E \frac{dN_\gamma}{dEdt} = 2 \frac{d\epsilon_a}{d\omega} \delta(E-\omega/2) V_{NS} \Delta t \Gamma_{a\gamma\gamma}. \end{equation} 
We define $\Delta t$ below.  Dividing by $1/(4\pi d^2)$ to derive a flux, we obtain
 
\begin{equation} E\frac{d\Phi}{dE} = 2 \frac{d\epsilon_a}{d\omega} \delta(E-\omega/2) \frac{V_{NS} \Delta t \Gamma_{a\gamma\gamma} }{4\pi d^2}. \label{eq:energyFlux}\end{equation}

We model the timescale of axion emission from a nuclear medium as the mean free time, which is the mean time $\Delta t$ between successive axion emissions in the nuclear medium.  This is appropriate as we are modeling the instantaneous emission of axions from neutron stars.  The emission rate $\Gamma_a$ is given by Raffelt~\cite{RaffeltBounds} as:

\begin{equation} \Gamma_a = \frac{g_{ann}^2}{8 M_N^2} \omega S_\sigma(\omega). \end{equation}
We compute the mean free time from the emission rate, 

\begin{equation} \Delta t = \frac{\hbar}{\left<\Gamma_a\right>_\omega}, \end{equation}
by considering the average over the axion energy range that we consider, denoted by $\left<\Gamma_a\right>_\omega$.  Thus we have:
\begin{equation} \Delta t = \frac{8 \hbar M_N^2}{g_{ann}^2} \frac{\int d\omega}{\int d\omega \ \omega S_\sigma(\omega;\mu,T)}  \label{eq:meanfreetime} \end{equation}




Evaluating equation~\eqref{eq:meanfreetime} provides a mean free time of $\Delta t = 23.2 \ {\rm s }\left(\frac{\rm eV}{m_a}\right)^2$ for $T=20$ MeV.  
This provides a timescale $\Delta t$ for the emission of axions that occurs instantaneously in the neutron star, as we assume here.

\noindent Upon simplification of equation~\eqref{eq:energyFlux}, we obtain
\begin{equation} E \frac{d\Phi}{dE} = 1.8 \times 10^{-2} \left(\frac{m_a}{\rm eV}\right)^5 \left(\frac{\Delta t}{23.2 \ {\rm s}}\right)\left(\frac{100 \ {\rm pc}}{d}\right)^2  \left(\frac{2E}{100 \ {\rm MeV}}\right)^4 \left(\frac{S_\sigma(2E)}{10^7 \ {\rm MeV}^2}\right) \ {\rm  cm}^{-2} {\rm s}^{-1}. \label{eq:sed}\end{equation}

If flux limits from neutron stars from \emph{Fermi} LAT are on the order of $10^{-9}$ cm$^{-2}$s$^{-1}$, we expect our data to be sensitive to $m_a \sim \mathcal{O}({0.01 \rm eV})$ (since $S_\sigma$ is of the order $10^7$ MeV$^2$, and so $(m_a/{\rm eV})^5$ must be of the order 10$^{-7}$, in order to preserve the equality).  Note the strong dependence on axion mass $(m_a/{\rm eV})^5$.   

We may consider the effect on axion mass limits due to variations in the model parameters.  In the model of neutron stars that we are considering~\cite{ruster2005phase}, we may consider $10 {\rm \ MeV} \leq T \leq 50$ MeV, and $9 \leq \mu/T \leq 11$~\cite{ruster2005phase}, and we plot curves in Figure~\ref{fig:spinStruct}.  We quantitatively consider the effect of variations in these parameters on the axion limits in Section~\ref{sec:Observations}.
Qualitatively, increasing (decreasing) the assumed $T$ would tend to shift the spectrum towards higher (lower) energies.  The model flux depends on $\omega^4 S_\sigma(\omega;\mu,T)$, which increases with $T$, but the timescale depends on $\left(\int d\omega \ \omega S_\sigma(\omega;\mu,T)\right)^{-1}$, which decreases with $T$.  Thus, a simple calculation finds that the limits on $m_a$ would be smaller for $T = 50$ MeV, and larger for $T=10$ MeV.  The order of magnitude of the limits would still be the same for these changes in temperature.  
Increasing the degeneracy parameter $\mu$ would tend to decrease the amplitude of the spin-structure function.  At $\mu/T = 11$, the limits would be larger, and at $\mu/T = 9$, the limits would be smaller.  Changing the $k$ parameter would not affect the limits substantially.

\subsection{Axion-Photon Conversion}\label{sec:agamma}

In principle, photon to axion conversions might take place in pulsar magnetospheres, as shown in detail in Ref.~\cite{perna}.  This process might compete with axion decays.  Yet, it turns out that it is a negligible effect for the case considered here, as described, e.g., in Ref.~\cite{raffelt}.  Specifically, the mixing angle is shown to be very small for axions with energies  $\sim$ 100 MeV and magnetic field strengths of order 10$^{12}$ G. We now show that this process is negligible.

 The Lagrangian for the coupling between the electromagnetic field and the axion field may be written as~\cite{axionBook}: 
 \begin{equation} \mathcal{L} = g_{a\gamma}\mathbf{E}\cdot\mathbf{B}a. \end{equation}

\noindent The mixing term is given by
\begin{equation} \Delta_{a\gamma} \simeq 0.98\times 10^{-9} {\rm eV} g_{10} B_{12} \end{equation}
 where 
\begin{equation} g_{10} = \frac{g_{a\gamma}}{10^{-10} \ {\rm GeV}^{-1}} \end{equation} 

\noindent and $B_{12} = B/(10^{12} \ {\rm G})$.  The probability of photon--axion mixing is proportional to $\sin^2(2\theta)$, where $\theta$ is the mixing angle; $\sin^2(2\theta)$, in the vicinity of pulsars, is given by

\begin{equation} \sin^2(2\theta) = \frac{4\Delta_{a\gamma}^2}{\Delta^2_\parallel + 4\Delta^2_{a\gamma}} = 4.5\times 10^{-16} g_{10}^2 B_{12}^{-2} \omega_{\rm MeV}^{-2}.\label{eq:Pagamma}
 \end{equation} 
\noindent From equation \eqref{eq:Pagamma}, since $\sin^2(2\theta)$ is small, the probability for conversion will be small, so that it is justified to completely ignore the axion--photon conversions in the pulsar magnetosphere. 
In equation \eqref{eq:Pagamma}, the QED vacuum birefringence term to first order is given by
\begin{equation} \Delta_{\parallel} = 0.92\times 10^{-1} {\rm eV} B_{12}^2 \omega_{\rm MeV}, \end{equation}
where $\omega_{\rm MeV}$ is the photon energy in MeV.  QED vacuum birefringence refers to the phenomenon that the parallel and perpendicular polarization states may have different refractive indices in vacuum.  The plasma term ($\Delta_{pl}=-\omega^2_{\rm pl}/(2\omega)$) for mixing can be neglected compared to  $\Delta_\parallel$~\cite{raffeltConversion}.\footnote{The plasma frequency is $\omega_{\rm pl}^2=4\pi\alpha n_e/m_e$, where the typical electron density $n_e$ in the vicinity of pulsars is on the order of $10^{11}$ cm$^{-3}$~\cite{GJ}.} In effect, the axion can be treated as massless in this formalism.

\begin{figure}[htb]
\begin{centering}
\includegraphics[width=6in]{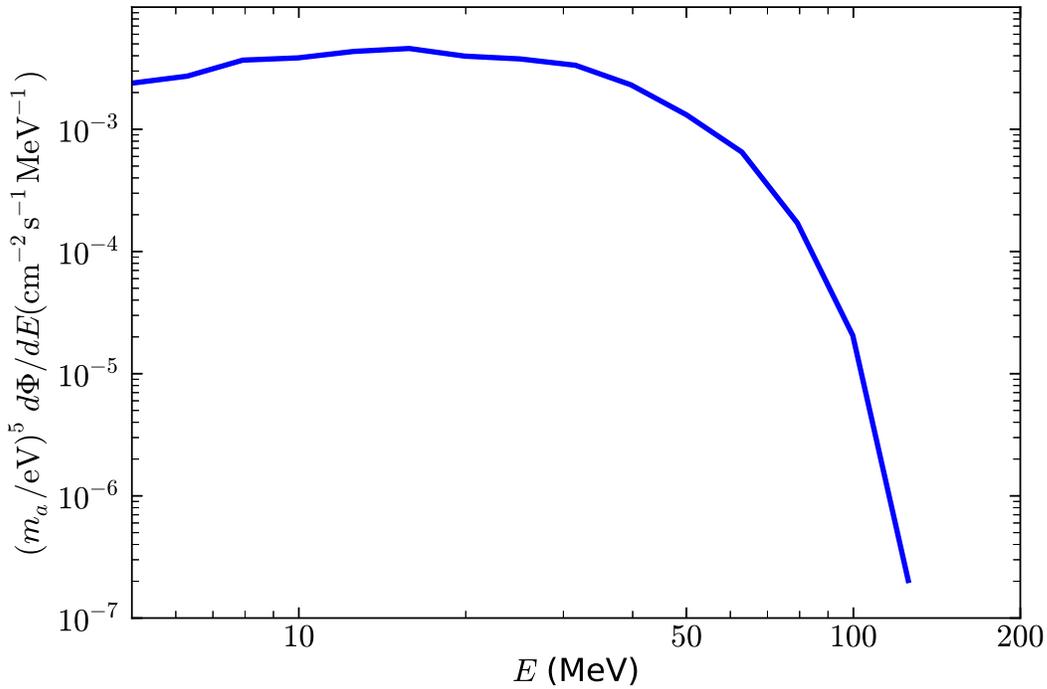}
\caption{The spectral energy distribution of gamma rays from axion decays, for the neutron star J0108--1431, derived according to equation \eqref{eq:sed}.}\label{fig:sed}
\end{centering}
\end{figure}

\section{Observations\label{sec:Observations}}

Four neutron stars were chosen from the most extensive pulsar catalog available, the ATNF catalog~\cite{psrcat}, to satisfy several criteria.  We require that the distance $d<0.4$ kpc, since the limits are degraded as $d^{-2}$, and since the nearest neutron star considered is at a distance of $d=0.24$ kpc.  Adding sources beyond 0.4 kpc is expected to provide marginal improvement to the combined limit on $m_a$.  We also require for the Galactic latitude that $|b|>15^\circ$, in order to avoid contamination from diffuse emission from the Galactic plane, which is significant at the low energies we consider here.  In addition, we require that there are no sources from the 2$^{nd}$ \emph{Fermi} LAT Catalog (2FGL) closer than $1.5^\circ$ away from the center of the region of interest (ROI) corresponding to each source, again, to limit contamination since the LAT point spread function (PSF) is broad at low energies, approximately 5$^\circ$ at 60 MeV for front converting events.  However, the PSF improves with increasing energy, to 3.5$^\circ$ at 100 MeV to 2$^\circ$ at 200 MeV.  The 1.5$^\circ$ cutoff was determined empirically, as it was noticed that sources farther than $1.5^\circ$ did not affect the test statistic corresponding to a null detection.  In particular, four sources that had 2FGL sources closer than 1.5$^\circ$, J1856-3754, J0030+0451, J1045-4509, J0826+2637, were rejected based on this criterion.  We do use the sources J0108-1431, J0953+0755, J0630-2834, J1136+1551, which are the only sources that satisfy these criteria.  The neutron star sources considered, and their physical parameters, are excerpted from the ATNF pulsar catalog \cite{psrcat}, and are listed in Table \ref{tab:sources}.  

A five-year data set corresponding to August 2008--August 2013 (MET 239557417--397323817), as obtained from the \emph{Fermi} Data Catalog, is extracted in a circular region of 20$^\circ$ radius about the coordinates for each neutron star.  In consideration of the spectral model of axions decaying into gamma rays, in addition to the \emph{Fermi} LAT Galactic diffuse model and the current LAT instrument response function, photons with energy 60 MeV--200 MeV were used.  The diffuse model is not computed below 56 MeV.  The axion model spectrum has a negligible contribution above 150 MeV, so it is not necessary to consider energies much above that, since the expected flux drops below the sensitivity of the LAT. 

We use data selected with the Source Event Class criteria, which is the recommended selection for the analysis of point sources.\footnote{The Source Event Class is described at \\
 \url{http://fermi.gsfc.nasa.gov/ssc/data/analysis/documentation/Cicerone/Cicerone_Data_Exploration/Data_preparation.html}}. We select on front-converting events, namely, events that convert in the front section (thin layers) of the tracker of the LAT, using \texttt{FTOOLS}~\cite{heasoft}.  The front events are known to have a narrower point spread function than the events converting in the back of the tracker~\cite{p7rep}.  
We used the \texttt{P7REP\_SOURCE} data with the corresponding instrument response function, \texttt{P7REP\_SOURCE\_V15::FRONT}.  The Galactic diffuse model and isotropic model used were appropriate to this instrument response function, and were  \emph{gll\_iem\_v05.fits}  and  \emph{iso\_source\_front\_v05.txt}, respectively.\footnote{These models may be obtained from \url{http://fermi.gsfc.nasa.gov/ssc/data/access/lat/BackgroundModels.html.}}  We performed an analysis where the likelihood function is unbinned with respect to energy, in order to model the likelihood function with each photon treated independently.    

We modeled background point sources according to the 2FGL catalog~\cite{2FGL}.  They were modeled with free normalizations and fixed spectral indices within 10$^\circ$ of the ROI center; outside of the 10$^\circ$ radius circle the normalizations and spectral indices were fixed.  No emission was detected from any of the 4 neutron star sources, and one--sided upper limits at the 95\% confidence level were placed individually for each neutron star source.  In addition, a combined upper limit from statistically combining the ROIs by joint likelihood analysis was obtained.

We model the putative signal as a normalization factor multiplied by the expected $d\Phi/dE$ for that source, corresponding to the axion decay flux spectrum.  When optimizing the likelihood function, the normalization is the only free parameter for the axion signal from the neutron star source.

\begin{table}
\begin{centering}
\begin{tabular}{|l|l|l|l|l|l|l|l|}
\hline
Source Name & RA ($^\circ$) & Dec.($^\circ$) & $\ell$ ($^\circ$) & $b$ ($^\circ$)   & $d$ (kpc) & Age (Myr) & $B_{\rm surf}$ (G) \\
\hline
J0108-1431 & 17.035 & -14.351 & 140.93 & -76.82 & $0.240^{+0.124}_{-0.061}$ & 166 & 2.52$\times10^{11}$ \\
\hline
J0953+0755 & 148.289 & 7.927 & 228.91 & 43.7 & $0.262^{+0.005}_{-0.005}$ & 17.5 & 2.44$\times10^{11}$ \\
\hline
J0630-2834 & 97.706 & -28.579 & 236.95 & -16.76 & $0.332^{+0.052}_{-0.040}$ & 2.77 & 3.01$\times10^{12}$ \\
\hline
J1136+1551 & 174.014 & 15.851 & 241.90 & 69.20 & $0.360^{+0.019}_{-0.019}$ & 5.04 & 2.13$\times10^{12}$ \\

\hline
\end{tabular}
\end{centering}
\caption{Table of sources and their coordinates, distances, ages, and surface magnetic field strengths.   Data excerpted from Ref.~\cite{psrcat}.  Distance uncertainties have been extracted from Refs.~\cite{pulsarParallaxes,pulsarJ1136,parallax2002}.}\label{tab:sources}

\end{table}
 
\begin{figure}
\begin{centering}
\includegraphics[width=3in]{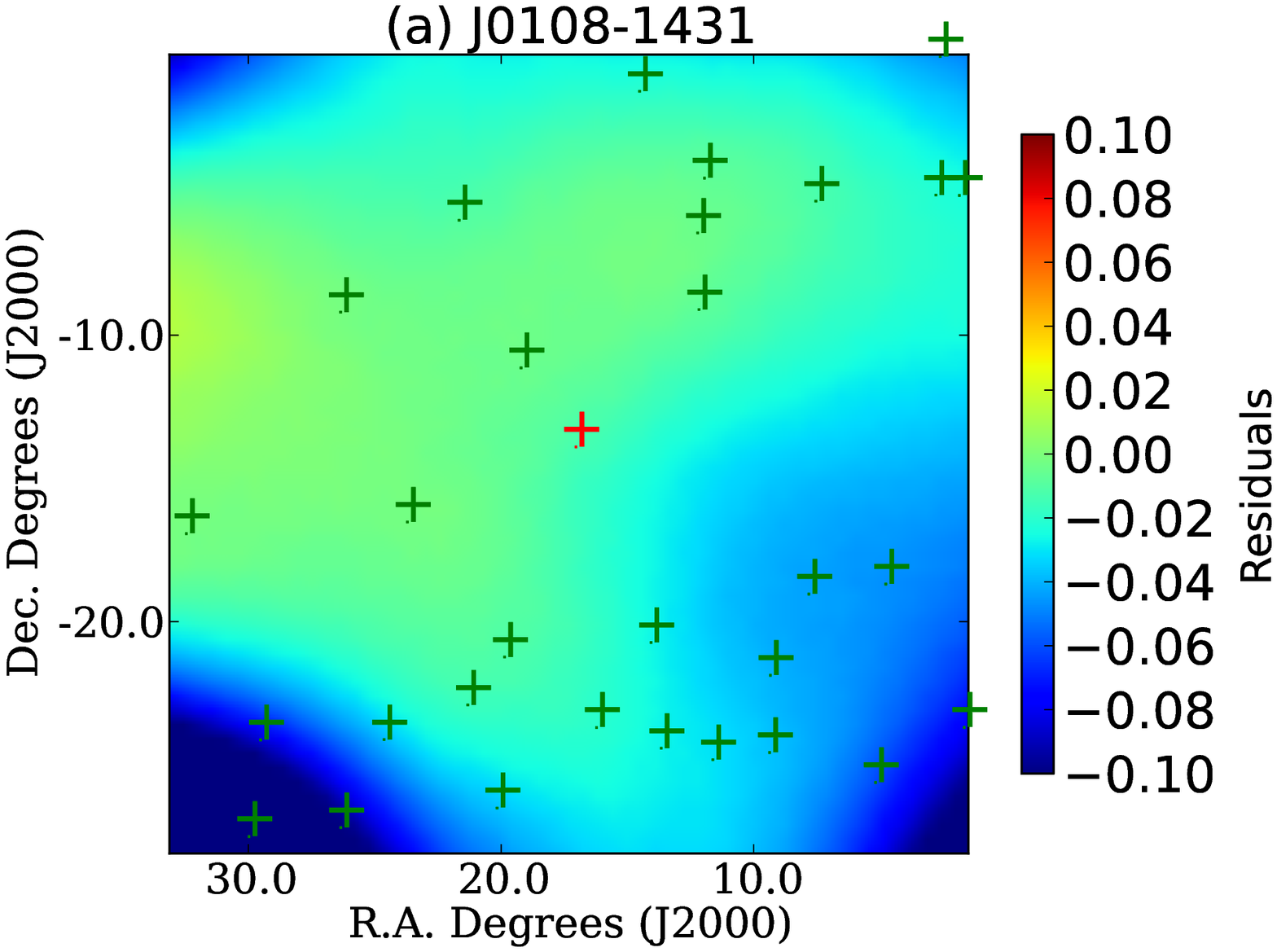}
\includegraphics[width=3in]{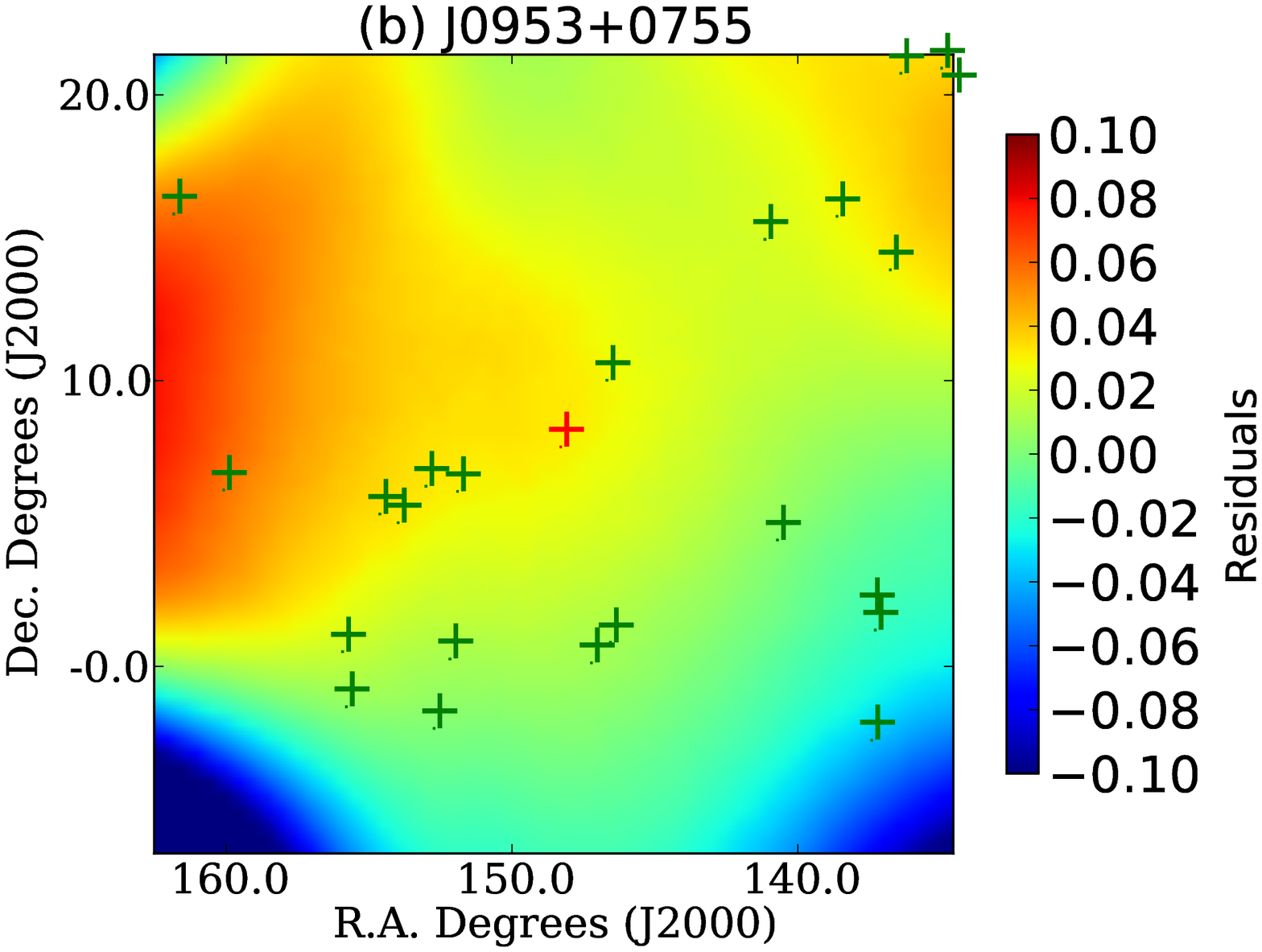} 
\includegraphics[width=3in]{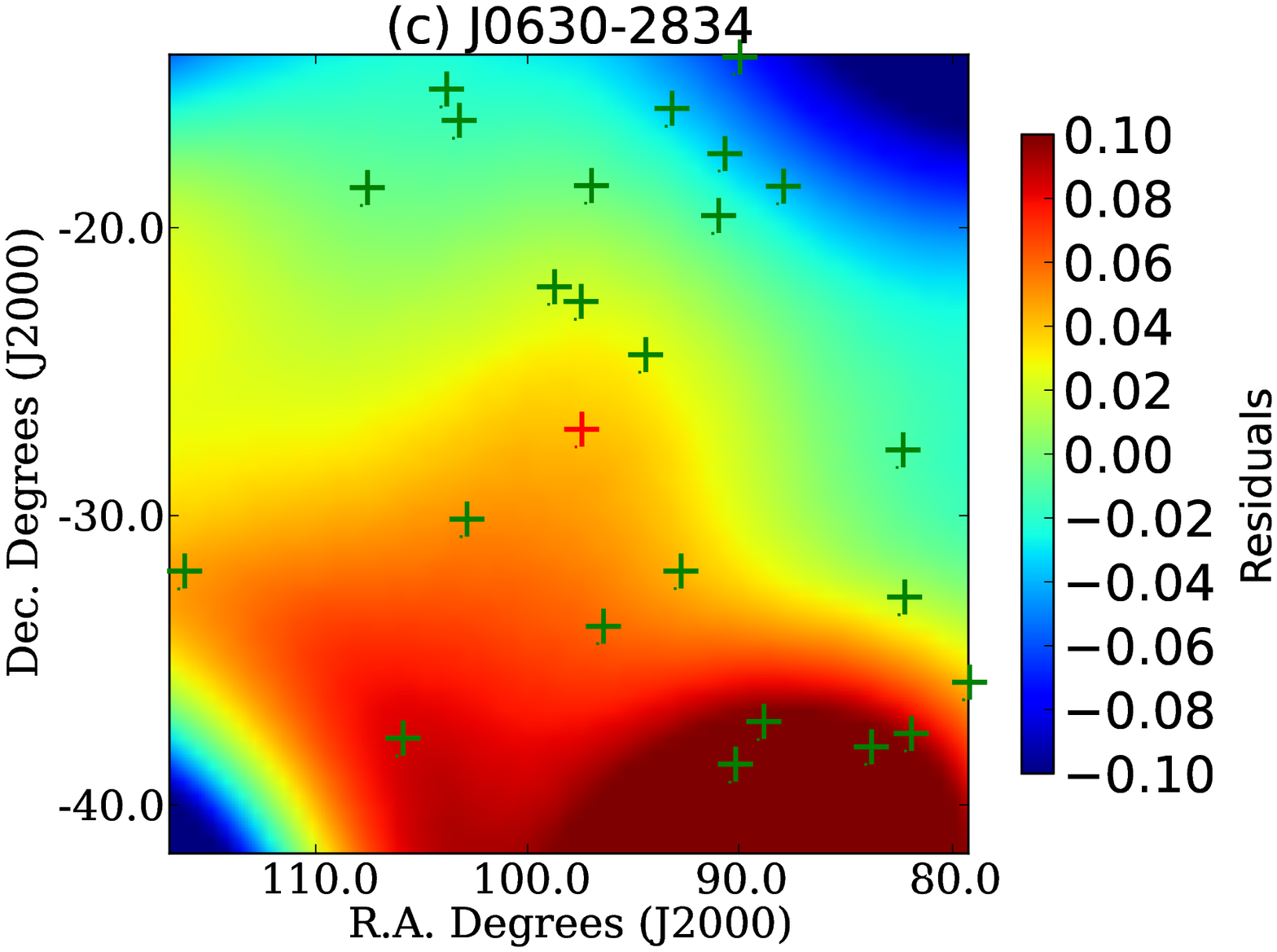}
\includegraphics[width=3in]{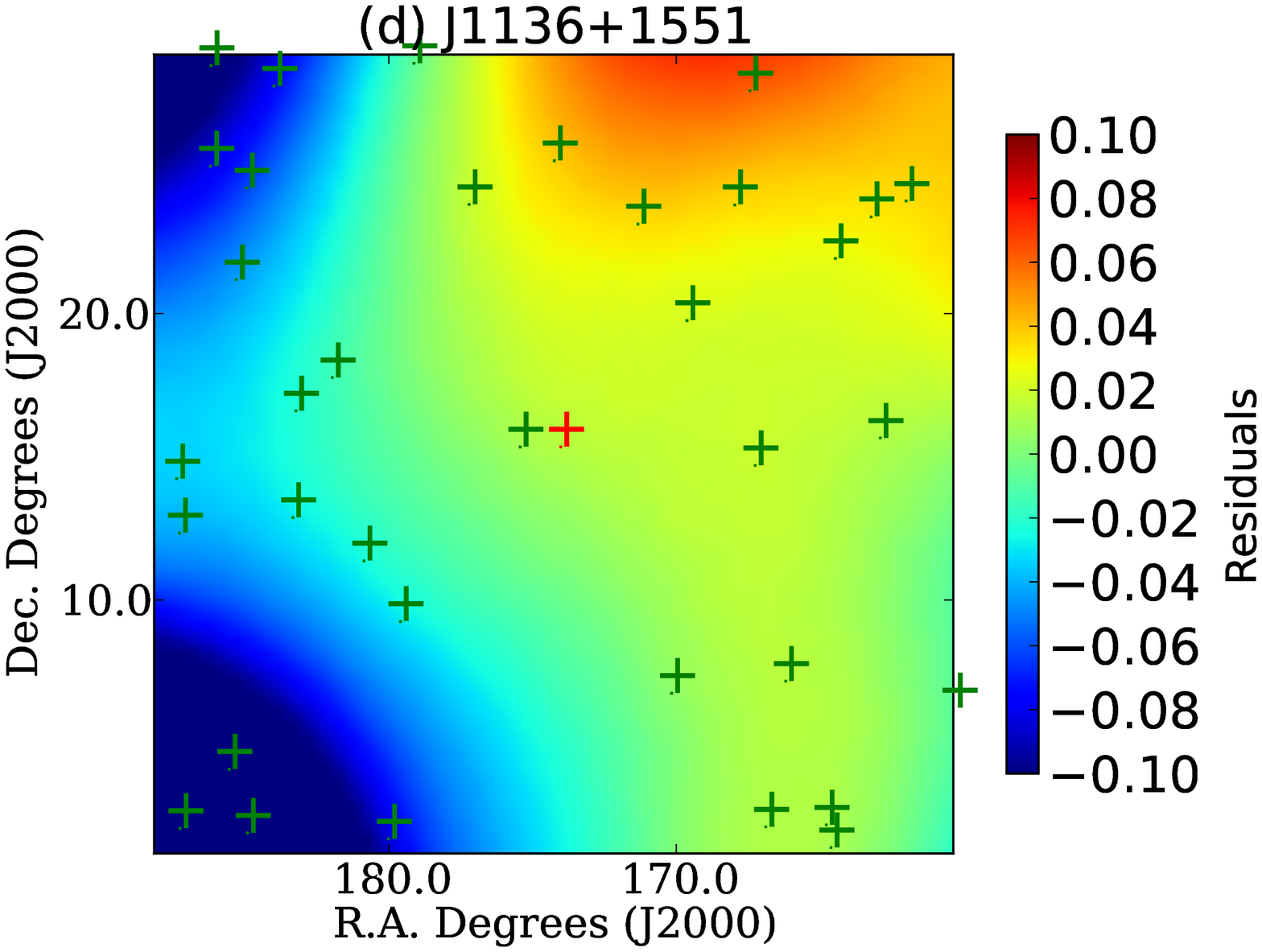} 
\caption{Smoothed residual maps, computed according to (counts-model)/model, in celestial coordinates for the sources (a) J0108-1431 (b) J0953+0755 (c) J0630-2834 (d) J1136+1551.  The energy range is 60-200 MeV.  The residual maps are plotted in a $28^\circ\times28^\circ$ square, with the neutron star at the center, with a pixel size of  0.5$^\circ$. The images are smoothed with a Gaussian of sigma = 5$^\circ$.  2FGL sources are denoted with green crosses, the putative neutron star source is denoted with a red cross in the center.}  \label{fig:residMap}
\end{centering}
\end{figure}

\begin{figure}
\begin{centering}

\includegraphics[width=\textwidth]{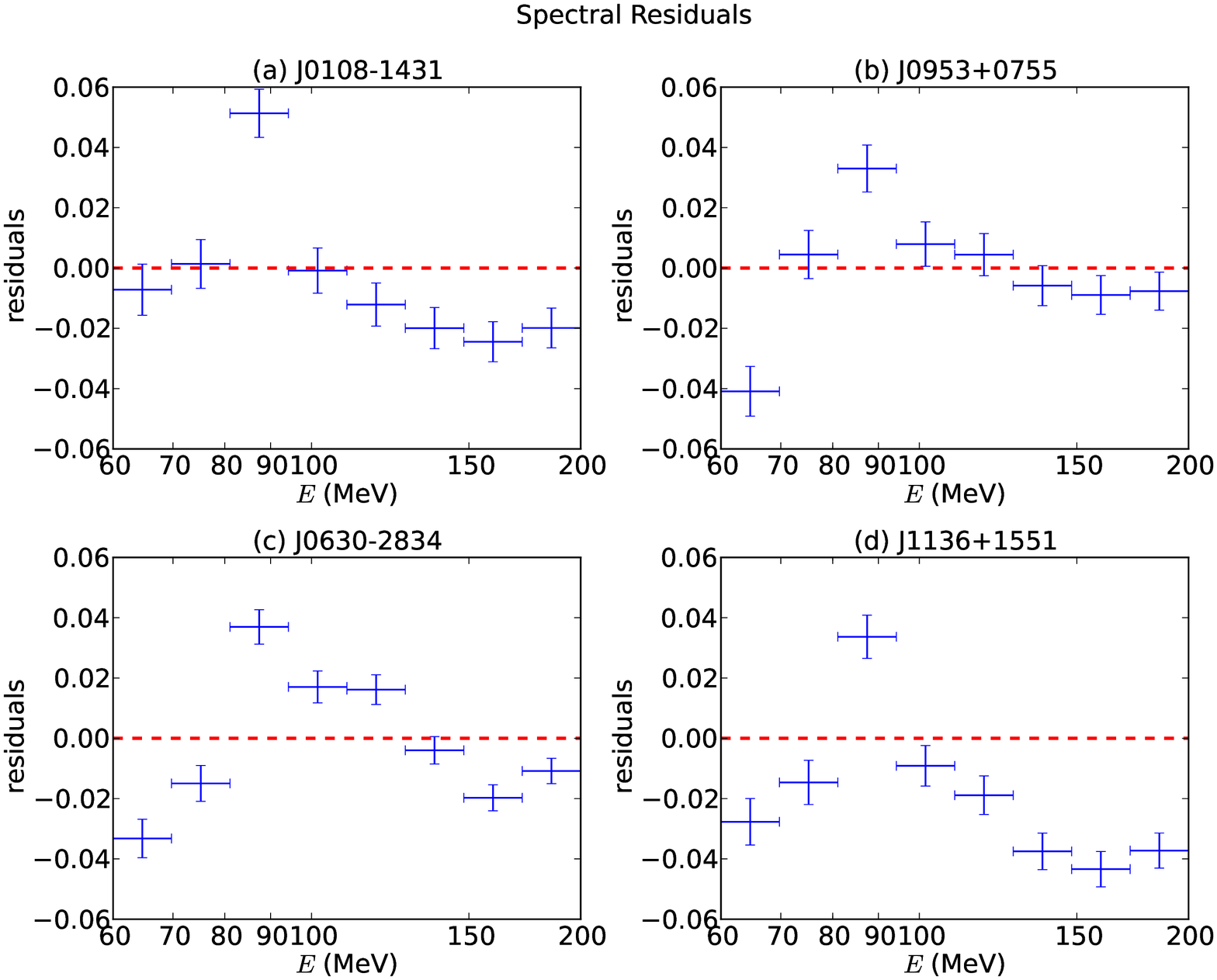}
\caption{Spectral residuals over the ROI, as a function of energy, according to (counts-model)/model, for the sources (a) J0108-1431 (b) J0953+0755 (c) J0630-2834 (d) J1136+1551.  The errorbars along the $x-$axis denote the range of energies considered at each point, and the errorbars along the $y-$axis are statistical uncertainties.}\label{fig:resid}
\end{centering}
\end{figure}

In Figure \ref{fig:residMap}, we show residual maps, for which the fractional difference between count and model is computed pixelwise, for the 4 sources.  The spectral residuals are at most 14\%.  In Figure \ref{fig:resid}, we plot the residuals quantifying the discrepancy between the spectral model and the actual data summed over the entire ROI.  The residuals are at most 6\% in the range of energies examined.  The best agreement occurs at the low and high end of the energy range considered.  It may be noticed across the four panels that the points near 90 MeV show significant positive deviations, which is also observed in many of the blank field samples.  This may be due to the systematic uncertainties from modeling the data at energies below 100 MeV with the Fermi-LAT. One possible explanation is that the model spectra for the 2FGL background point sources are not accurate below 100 MeV, since they were fit above 100 MeV; the 2FGL sources are modeled by power-law functions. Another possible explanation lies in modeling the diffuse emission at these energies; see Section~\ref{sec:Discussion}.  Furthermore, the spatial residuals near 90 MeV are not consistent with coming from a point source at the positions of the neutron stars.

\subsection{Upper Limits on the Axion Mass}\label{sec:ul_ma}

Using the Fermi--LAT \emph{ScienceTools}, we compute one-sided upper limits using MINOS~\cite{minuit}.  We first find the maximum of the likelihood function $\mathcal{L}$, and calculate where the $2\Delta\log \mathcal{L}$ function has increased by 2.71.  This corresponds to limits at the 95\% confidence level.  We take the upper limit on the normalization parameter, and consider the upper limit on the axion mass as the normalization to the power of 1/5, since all the other astrophysical dependences are already considered in $d\Phi/dE$.  The test statistic (log-likelihood ratio test between the hypothesis of no source versus a source) for all 4 sources is consistent with 0, within numerical precision, indicating a null detection.  The results are shown in Table \ref{tab:ul_ma}.  The combined upper limits, computed using \emph{Composite2} module of the \emph{ScienceTools}, are somewhat more stringent than those for the source yielding the best limits, J0108-1431. The composite likelihood analysis sums the likelihood functions corresponding to the individual ROIs, and in this analysis, obtains the normalization parameter corresponding to the model spectrum as a tied parameter over the four ROIs. 

The systematic errors have three main components.  Since the spectral residuals are of order 5\%, 5\% of the diffuse flux within a 1 PSF radius ($\sim 5^\circ$)  of each source is added to the flux upper limit.\footnote{5$^\circ$ is the 68\% containment of the PSF corresponding to 60 MeV energies for front events.}  The mean diffuse flux computed from the 4 ROIs is 2.63$\times 10^{-9}$ cm$^{-2}$ s$^{-1}$.  Since the systematic error related to the instrument response function of the LAT is estimated to be 20\% of the flux at these energies~\cite{p7rep}, we revise the flux estimate upwards by 20\%.  From the uncertainties on the neutron star distances $\delta d$, we propagate the relative uncertainties from the neutron star distances on the axion mass as $2\delta d/d$, which is as high as 103\% for source J0108-1431.  These three sources of systematic errors were added linearly, in addition to the systematic uncertainty arising from the theory (see below).
As the flux is proportional to the normalization parameter, we may apply systematic corrections on the flux to compute the limits on the axion mass. 

In Table~\ref{tab:ul_ma_model}, we present the limits for the different model parameters of $T$ and $\mu/T$ discussed in Section~\ref{sec:Pheno} for the source J0108-1431.  We observe that for $T=10$ MeV, the limits on $m_a$ are less stringent, and they are more stringent for $T=50$ MeV.   Also, for $T=20$ MeV, the smaller the $\mu/T$ parameter, the more stringent the limits.
The upper limits for the $T=10$ MeV, $\mu/T=10$ case are equal to the $T=20$ MeV, $\mu/T=10$ case.  The largest uncertainty from the parameters corresponding to the different models arises from the $T=20$ MeV, $\mu/T=11$ case, which has a $m_a$ of 42\% larger as compared to the reference model, the $T=20$ MeV, $\mu/T=10$ case.  The upper limit taking into account the systematic from the theory is $m_a < 7.9\times 10^{-2}$ eV.

 \begin{table}
\begin{centering}
\begin{tabular}{|l|l|l|l|l|l|}
\hline
 & J0108-1431 & J0953+0755 & J0630-2834 & J1136+1551 & Combined \\
\hline
95\% CL u.l.  & 6.4 &  9.4 & 10.2 & 10.9 &  7.9  \\ 
for $m_a$ ($10^{-2}$ eV) & & & & & \\
\hline
95\% CL u.l.  & 4.03  & 7.40 & 4.82 & 8.52 & - \\
on the flux & & &  &  &\\
 (10$^{-9}$ cm$^{-2}$ s$^{-1}$) & & & & &\\

\hline
\end{tabular}
\end{centering}
\caption{Table of 95\% Confidence Level upper limits on $m_a$ and the flux, for the various sources taken individually, as well as the combined limit.  The flux and mass upper limits have been corrected for systematic uncertainties, including uncertainties from theory.  In addition, the limits on $m_a$ account for the uncertainties on the neutron star distances as described in Section \ref{sec:ul_ma}, as well as the systematic uncertainty from the theory described in Table~\ref{tab:ul_ma_model}.}\label{tab:ul_ma}
\end{table}

\begin{table}
\begin{centering}
\begin{tabular}{|l|l|l|l|l|l|}
\hline
 & $T=10, \ \mu/T = 10$ & $T=20, \ \mu/T = 9$ & $T=20, \mu/T = 10$ & $T=20, \mu/T = 11$ & $T=50, \ \mu/T = 10$ \\
\hline
95\% CL u.l.  & 6.4 &  4.1 & 6.4 & 9.1 &  4.9 \\ 
for $m_a$ ($10^{-2}$ eV) 
 & & & & & \\
\hline
95\% CL u.l.  &  3.94 & 4.26 & 4.03 & 3.98 & 4.30 \\
on the flux & & &  &  &\\
 (10$^{-9}$ cm$^{-2}$ s$^{-1}$) & & & & &\\

\hline
\end{tabular}
\end{centering}
\caption{Table of 95\% Confidence Level upper limits on $m_a$ and the flux, for different model parameters, for the the source J0108-1431.  The flux and mass upper limits have been corrected for systematic uncertainties.  In addition, the limits on $m_a$ account for the uncertainties on the neutron star distances as described in Section \ref{sec:ul_ma}.  We also present the percent change in the upper limit on $m_a$ from the $T=20, \mu/T = 10$ reference model.  We obtain a systematic uncertainty of 42\% from the theory. }\label{tab:ul_ma_model}
\end{table}

\begin{figure}
\begin{centering}
\includegraphics[width=6in]{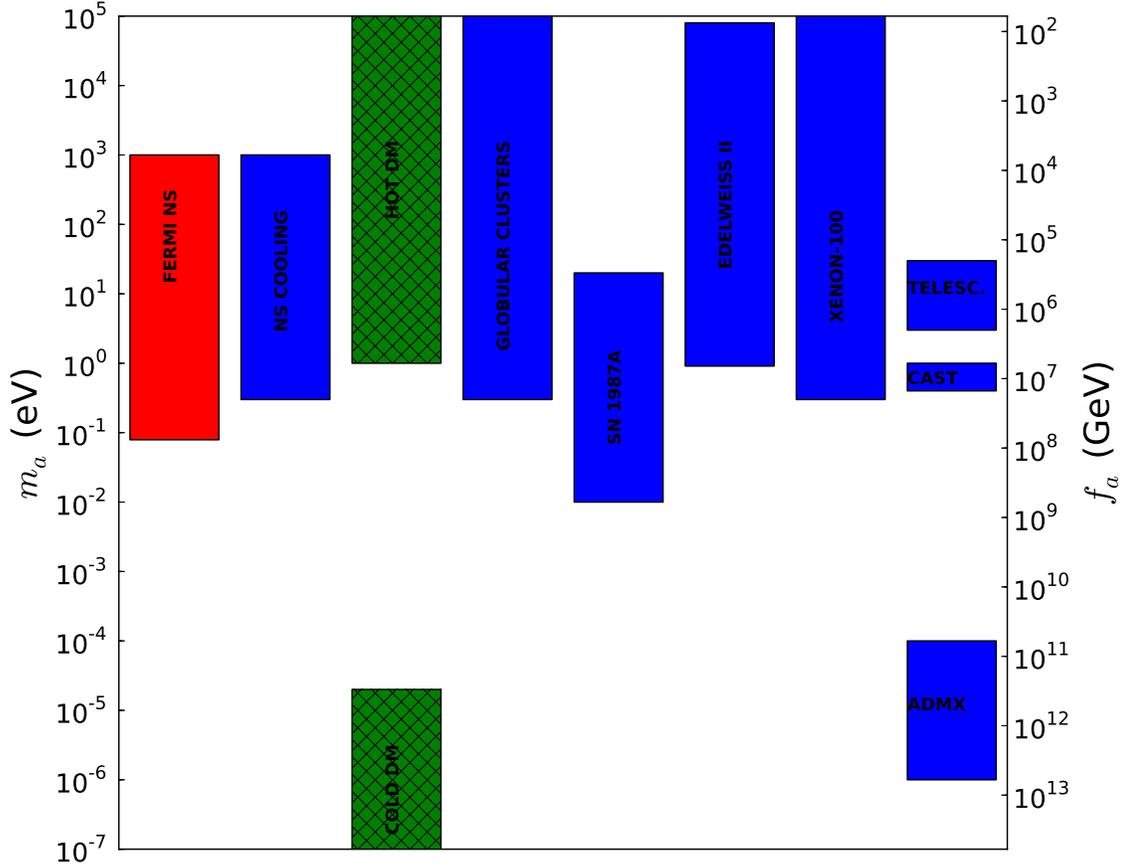}
\caption{Exclusion ranges for $m_a$ and $f_a$ derived here (red), compared with those of other experiments (blue)~\cite{PDG,EdelII,Xenon100} and theoretically motivated inclusion regions (green)~\cite{coldDM,hotDM1,hotDM2}; values within the shaded boxes (blue and red) are excluded, with the lower bound corresponding to the 95\% CL upper limit.}\label{fig:exclRange}
\end{centering}
\end{figure}

In Figure ~\ref{fig:exclRange}, we show the excluded region of $m_a$ from this analysis as compared to that corresponding to other astrophysical studies.  We may note that the exclusion region $m_a > 7.9\times 10^{-2}$ eV is valid until $m_a \simeq 1$ keV; for heavier axions, the assumption of relativistic axions is no longer valid. 

\subsection{Axion--Like Particles}
\begin{figure}
\begin{centering}
\includegraphics[width=6in]{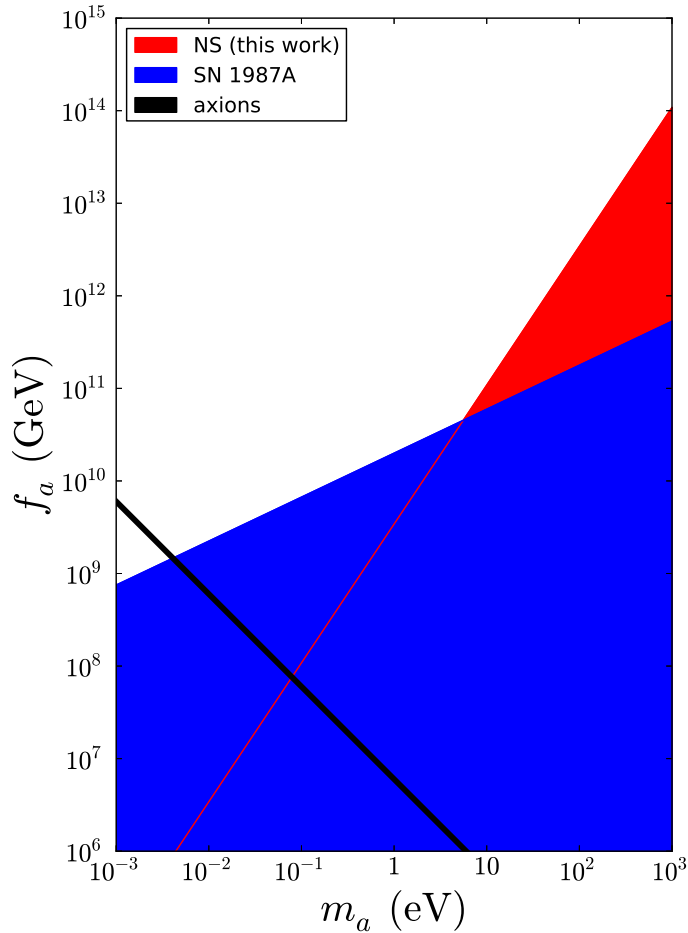}
\caption{Exclusion plot for the $(m_a,f_{a})$ parameter space for ALPs at 95\% CL\@.  NS describes the region derived from this work, SN 1987A describes the region derived from \emph{Fermi} LAT analysis of SN 1987A \cite{giannotti}. The axion line (black) shows the parameters allowed by PQ-axions.}\label{fig:exclPlot}
\end{centering}
\end{figure}

We generalize the results in this analysis to consider ALPs as well.  By relaxing the criteria in equation \eqref{eq:mafa},  we obtain for the lifetime the relation: 

\begin{equation} \tau_{a\to\gamma\gamma} = 1.7\times 10^{17} f_{12}^2 \left(\frac{m_a}{{\rm MeV}}\right)^{-3} \ {\rm s} \end{equation}

\noindent where $f_{12}=f_a/10^{12}$ GeV.  In addition, the axion-nucleon coupling can be expressed in terms of $f_a$, which is more fundamental, as~\cite{axionBook}:

\begin{equation} g_{\rm ann} = (M_N/f_a)c_N \simeq \frac{0.939 \ {\rm GeV}}{f_a} 0.1 \simeq \frac{10^{-13}}{f_{12}} \end{equation}

\noindent where $c_N$ was introduced in Section \ref{sec:Pheno}.
\noindent We now express the energy flux in terms of $f_{12}$ and $m_a$ as follows:

\begin{equation} E \frac{d\Phi}{dE} =  6.48\times 10^{-13} f_{12}^{-2} \left(\frac{m_a}{\rm eV}\right)^3 \left(\frac{\Delta t}{23.2 \ {\rm s}}\right) \left(\frac{100 \ {\rm pc}}{d}\right)^2 \left(\frac{2E}{\rm MeV}\right)^4  \left(\frac{S_\sigma(2E)}{10^7 {\rm MeV}^2}\right) \ {\rm cm}^{-2} {\rm s}^{-1}. \end{equation}
Based on our upper limits, we exclude regions in the $(m_a,f_a)$ parameter space, as shown in Figure~\ref{fig:exclPlot}.
The region derived from analyzing neutron stars with \emph{Fermi}--LAT data excludes a larger portion of the parameter space above 1 eV than the region derived from analyzing SN 1987A as in Ref.~\cite{giannotti}.  This can be accounted for by the different dependence on the model parameters: this model depends on $m_a^3 f_a^{-2}$, whereas other models, such as described by Giannotti et al.~\cite{giannotti}, depend on $m_a^2 f_a^{-4}$.  The limits provided from this analysis are complementary to the other astrophysical limits, as we are examining a different physical process for axion emission. 

\subsection{Blank Fields}

\begin{figure}
\begin{centering}
\includegraphics[width=6in]{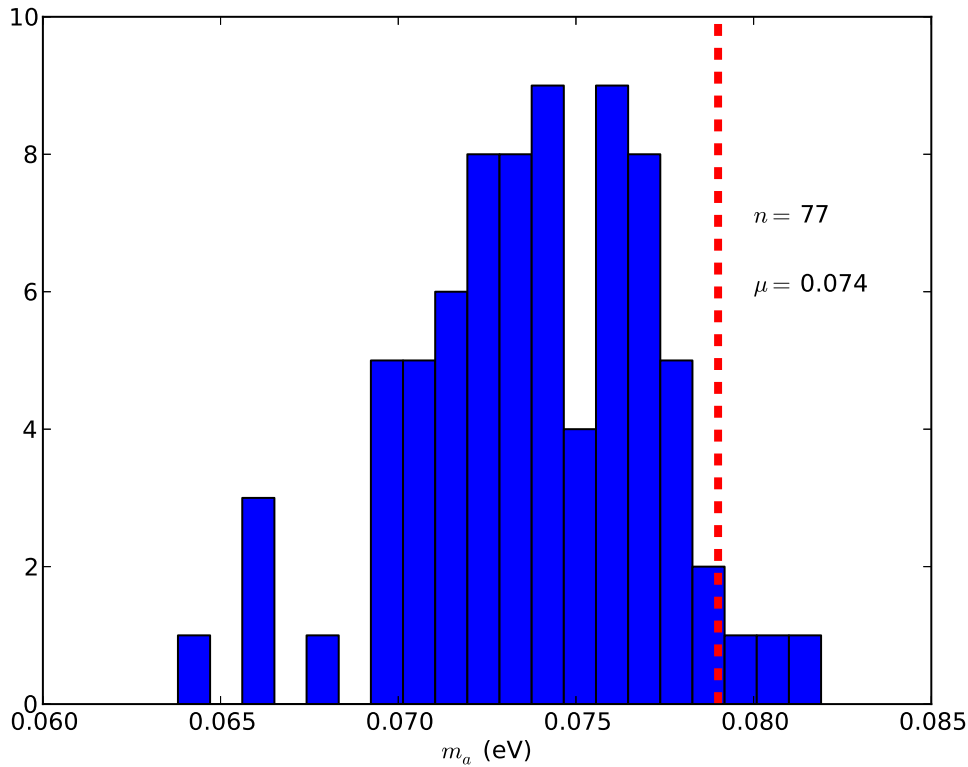}
\caption{A histogram of limits on $m_a$ from the joint likelihood analysis of the blank fields considered.  The red dashed line corresponds to the upper limit of 0.079 eV obtained from the analysis of the four selected neutron stars.}\label{fig:blankFields}
\end{centering}
\end{figure}

In order to validate the upper limits on $m_a$ from neutron star sources, we considered obtaining limits from blank regions of the sky.  We consider 25 high-latitude ROIs ($b>45^\circ$) distributed randomly over the sky, centered on an imaginary source.  A similar technique has been described in Ref.~\cite{dwarfPaper}.    We consider 77 random combinations of 4 ROIs randomly drawn from the sample of 25 ROIs,  and the limits for four ROIs are evaluated at the distances and magnetic fields of J0108-1431, J0953+0755, J0630-2834, and J1136+1551.  The joint likelihood is calculated from the 4 randomly drawn ROIs to obtain an upper limit.  The upper limits are revised upwards to account for the systematic uncertainties, in keeping with the procedure for the upper limits from data.  In Figure ~\ref{fig:blankFields}, we histogram the 77 limits on $m_a$.  The mean of the distribution is 0.077 eV, while the range is between 0.065 eV and 0.082 eV.  The combined limit for the four targets we evaluated, of 7.9$\times 10^{-2}$ eV, is slightly above the mean, but consistent with the blank field limit distribution.  

\section{Estimation of the Systematic Uncertainties due to Energy Dispersion with Simulations}\label{sec:sim}

In our fitting of the data and placing upper limits in earlier sections of the paper, we did not consider energy dispersion, because the analysis is unbinned.  Energy dispersion accounts for the difference between measured and true photon energy in the LAT.  It is important to consider energy dispersion because the effective area is changing rapidly and the spectra of the sources are strongly dependent on energy.  By considering energy dispersion in our simulation, and fitting with and without energy dispersion for the point sources, we may obtain an estimate of the systematic error from not taking into account the energy dispersion in our analysis. As the first step of this procedure, a simulation of one of the source regions is performed, namely, J0108-1431, as this ROI provides the best limits in the data analysis. Further details of the simulation are provided in Appendix~\ref{appB}. As the second step of the procedure, fitting of the simulated data according to the astrophysical model, within a specified ROI and a specified energy range (see below), is carried out.  
Finally, upper limits on the flux are derived for the putative source from gamma--rays arising from the axion spectral model, and systematic errors are computed.

The analysis of the simulated files is otherwise the same as that of the LAT data, except for using the binned analysis steps.   Photons within the 20$^\circ$ radius ROI were used in the fitting.  We perform a binned analysis in order to fit with energy dispersion.  In the case of the fitting of the data, a fit with energy dispersion is not performed because it is too computationally expensive in an unbinned analysis.  The \emph{ScienceTools} have a feature which allows for energy dispersion to be turned either on or off for the various components of the model.  Using this feature, fitting is performed in two cases: a) energy dispersion enabled for the point sources; b) energy dispersion disabled for the point sources.  In both cases a) and b), energy dispersion is disabled for the diffuse components.  The diffuse models are data--driven, and thus in a fit to real data, energy dispersion should be disabled for the diffuse models. We are repeating the same analysis as with data for the point source corresponding to J0108-1431.

We determine the systematic error on the fit without energy dispersion for the point sources by comparing to the fit with energy dispersion for the point sources, in other words, by comparing case a) with case b).  In Table~\ref{tab:simUpperLim}, we compare the flux upper limits for case a) and case b).  By considering the ratio of the flux upper limits between cases a) and b), the systematic error is estimated to be $\sim$5\%, which is in line with expectations about the LAT instrument performance at low energies~\cite{p7rep}.  We conclude that the upper limit on the flux is lower in the case of energy dispersion because the photons contributing to the upper limit have been shifted to higher or lower energies, i.e., outside the analysis region.  

\begin{table}
\begin{centering}
\begin{tabular}{|l|l|}
\hline
case & flux u.l. (95\% CL) (10$^{-9}$ cm$^{-2}$s$^{-1}$)\\
\hline
(a) enabled energy dispersion & 4.00 \\
\hline
(b) disabled energy dispersion & 4.20 \\
\hline
\end{tabular}
\caption{Flux upper limits in fitting cases a) and b).  a) corresponds to enabled energy dispersion in fitting the point sources, b) corresponds to disabled energy dispersion in fitting the point sources.  In both cases a) and b), the simulations were created with energy dispersion enabled and all fits were performed with energy dispersion disabled for the diffuse emission.}
\label{tab:simUpperLim}
\end{centering}
\end{table}

\section{Discussion\label{sec:Discussion}}

We have presented a new spectral model for gamma rays from decays of axions and ALPs, and we have derived limits from the analysis of the data.   
The combined upper limit on $m_a$ of 7.9$\times 10^{-2}$ eV according to point source emission from axion decay corresponds to a lower limit on $f_a$ of 7.6$\times 10^7$ GeV.

As can be seen in Figure~\ref{fig:exclRange}, we exclude the higher end of the mass range for axions.  It is important to note that we are comparing the limits derived here with those probing different processes and mechanisms.  Umeda et al. in Ref.~\cite{iwamoto1}, cite an upper limit of 0.3 eV from neutron star cooling, and their model dependence, from the same emission process but a different emissivity calculation, is proportional to $m_a^2 T^4$.   We also exclude larger regions of parameter space than EDELWEISS, XENON, CAST, and limits from globular clusters.  EDELWEISS-II and XENON100 are direct detection dark matter experiments.  XENON100 relies on the coupling of axions or ALPs to electrons in deriving the limits~\cite{Xenon100}.  EDELWEISS-II probes axion production by $^{57}$Fe nuclear magnetic transition; thus, $g_{\rm ann}$ has a different model dependence than presented here~\cite{EdelII}.  The telescope (TELESC.) region is excluded by the non--observation of photons that could be related to the relic axion decay ($a\to\gamma\gamma$) in the spectrum of galaxies and the extragalactic background light~\cite{telesc1,telesc2}.  The Axion Dark Matter Experiment (ADMX) limits are probed by means of a haloscope and rely on cosmology and the axion being a dark matter component~\cite{admx0,admx}.   From inspection of the theoretically motivated regions, the results derived here are not consistent with axions as hot dark matter (Hot DM), but are consistent with axions as cold dark matter (Cold DM), the bounds of which are derived from cosmology~\cite{coldDM,hotDM1,hotDM2}.  The excluded region from SN 1987A derives from the burst duration expected from an axion energy loss compared with what was observed in neutrinos.  Although the SN 1987A result probes the nucleon--nucleon bremsstrahlung channel for axion production, as examined here, it excludes lighter axions than the result presented here: the exclusion region represents the tradeoff between heavier axions being more efficiently produced and lighter axions not being trapped within the core~\cite{SNbounds}.  The bounds from globular clusters (GC) derive from energy loss in the axion channel from stellar cores~\cite{GC}, thus probing the nucleon--nucleon bremsstrahlung process as was done here.  The CERN Axion Solar Telescope (CAST) probes solar axions converting into photons in the strong magnetic field of a helioscope~\cite{CAST}, and provides strong constraints.  

As shown in Figure~\ref{fig:exclPlot}, we exclude a larger region of the ALP ($m_a,f_a$) parameter space for $m_a \gtrsim 1$ eV than a previous study of the SN 1987A supernova remnant.  This is due to our model and the improved limits associated with the analysis.  The slope of $\log(f_a)$ versus $\log(m_a)$ is 3/2 rather than 1/2 in the case of the model by Giannotti et al.~\cite{giannotti}.  This is due to the different dependence on $m_a$ and $f_a$ in our model.  For $f_a=10^{12}$ GeV, we allow $m_a \lesssim 10$ eV, whereas Giannotti's result allows $m_a \lesssim 10^{3}$ eV.  Our results imply that ALPs produced from neutron stars should be light.

We imagine several refinements to the analysis that could be made in future work.  First and foremost the new event-level analysis (Pass 8), recently made available by the Fermi-LAT collaboration, might conceivably allow extending the analysis down to even lower energies (e.g., 30 MeV) thanks to the significantly larger acceptance.  In addition to that, the new PSF and energy-dispersion event types introduced in Pass 8 offer the possibility of selecting sub-samples of events with significantly better energy and/or angular resolution.  Finally, a better spectral and morphological modeling of the diffuse emission and a better spectral characterization of the ROIs, both of which will be available with the forthcoming fourth LAT source catalog (4FGL), will provide additional space to improve on the analysis.

\section{Acknowledgments}

The authors would like to thank the anonymous referee.

BB acknowledges support from California State University, Los Angeles as Lecturer (Adjunct Professor) in the Department of Physics and Astronomy in the College of Natural and Social Sciences.   JG acknowledges support from NASA through Einstein Postdoctoral Fellowship grant PF1-120089 awarded by the Chandra X-ray Center, which is operated by the Smithsonian Astrophysical Observatory for NASA under contract NAS8-03060, and from a Marie Curie International Incoming Fellowship in the project ``IGMultiWave'' (PIIF-GA-2013-628997).  MM acknowledges support from the Knut and Alice Wallenberg Foundation, PI: Jan Conrad.  

The \textit{Fermi} LAT Collaboration acknowledges generous ongoing support
from a number of agencies and institutes that have supported both the
development and the operation of the LAT as well as scientific data analysis.
These include the National Aeronautics and Space Administration and the
Department of Energy in the United States, the Commissariat \`a l'Energie Atomique
and the Centre National de la Recherche Scientifique / Institut National de Physique
Nucl\'eaire et de Physique des Particules in France, the Agenzia Spaziale Italiana
and the Istituto Nazionale di Fisica Nucleare in Italy, the Ministry of Education,
Culture, Sports, Science and Technology (MEXT), High Energy Accelerator Research
Organization (KEK) and Japan Aerospace Exploration Agency (JAXA) in Japan, and
the K.~A.~Wallenberg Foundation, the Swedish Research Council and the
Swedish National Space Board in Sweden.
 
Additional support for science analysis during the operations phase is gratefully
acknowledged from the Istituto Nazionale di Astrofisica in Italy and the Centre National d'\'Etudes Spatiales in France.

\appendix
\section{Analytic Simplification of the Phase Space Integrals}
\label{appA}

We describe briefly how the multi--dimensional phase space integral in equation \eqref{eq:spinStruct} is analytically simplified before numerical methods are applied.
In Ref.~\cite{raffeltBrems},  it is noted that 
\begin{equation} E_1 + E_2 - E_3 - E_4 + \omega = \frac{-2 p_3^2 - 2\vec{p}_1\cdot\vec{p}_2 + 2\vec{p}_1\cdot\vec{p}_3 + 2\vec{p}_2\cdot\vec{p}_3}{2M_N} + \omega, \end{equation}
after integrating out $p_4$ owing to the momentum $\delta$--function.
Letting the $\vec{p}_1$ define the $z-$direction, we use polar coordinates with $\alpha$ and $\beta$ the polar and azimuthal angles of $\vec{p}_2 $ relative to $\vec{p}_1$, and likewise $\theta$ and $\phi$ for $\vec{p}_3$.
We write the dot products between the momentum vectors as shown below:
\begin{equation} \vec{p}_1 \cdot \vec{p}_2 = p_1 p_2 \cos \alpha, \end{equation}
\begin{equation} \vec{p}_1 \cdot \vec{p}_3 = p_1 p_3 \cos \theta, \end{equation}
\begin{equation} \vec{p}_2 \cdot \vec{p}_3 = p_2 p_3 \cos \alpha \cos\theta +\sin\alpha\sin\theta\cos\beta, \end{equation}
\noindent We define $f(\beta)\equiv E_1+E_2-E_3-E_4+\omega$. 
We integrate the $\delta$--function over $d\beta$.
\begin{equation} \int_0^{2\pi} d\beta \delta[f(\beta)] = \frac{2}{\left|df/d\beta\right|_{\beta = \beta_1}}\Theta\left(\left|\frac{df}{d\beta}\right|_{\beta=\beta_1}^2\right) \end{equation}
where $\beta_1$ represents the root in the interval $[0,\pi]$.
The derivative may be expressed in the form
\begin{equation} \left|\frac{df}{d\beta}\right| = \sqrt{az^2 + bz + c} \end{equation}
where $z = \cos\alpha$,\ with the following definitions
\begin{equation} a = p_2^2(-p_1^2 - p_3^2 + 2p_1p_3\cos\theta), \end{equation}
\begin{equation} b = 2\omega M_N p_1 p_2 - 2p_1p_2p_3^2 - 2\omega M_N p_2 p_3\cos\theta + 2p_1^2p_2p_3\cos\theta + 2 p_2 p_3^3\cos\theta - 2p_1p_2p_3^2\cos^2\theta, \end{equation}
\begin{equation} c = \omega^2 M_N^2 + 2\omega M_N p_3^2 + p_2^2p_3^2 - p_3^4 - 2 \omega M_N p_1 p_3 \cos\theta + 2 p_1 p_3^3\cos\theta - p_1^2p_3^2\cos^2\theta - p_2^2p_3^2\cos^2 \theta. \end{equation}
After these analytic simplifications, we integrate the phase space integral with respect to $dp_1dp_2dp_3d\cos\theta d\cos\alpha$.  

\section{Details of the Simulation to Estimate Systematic Uncertainties Due to Energy Dispersion}
\label{appB}
We present the details of the simulation to estimate the systematic uncertainties.
Within a 20$^\circ$ radius ROI of source J0108-1431, all 2FGL point sources are simulated, as well as the same diffuse isotropic and galactic sources used in the fitting of the data.  The source J0108-1431 is not simulated in order to test our ability to set limits on a putative source. The \emph{ScienceTools} simulator \emph{gtobssim} is used to simulate photon events from astrophysical sources and to process those photons according to the specified instrument response functions, as discussed in Ref.~\cite{razzano}.  The 5 year FT2 (spacecraft) file in the simulation corresponds to the same file used in the analysis of J0108-1431 data. The P7REP\_SOURCE\_V15::FRONT instrument response function is used in the simulation.   We implement the simulation with energy dispersion by generating photons between 10 MeV to 400 MeV; this range is chosen to be larger than the fitting energy range (to be discussed below) due to energy dispersion. The isotropic diffuse source was extrapolated below 56 MeV to an energy of 10 MeV in order to accurately model the ROI at low energies.  The galactic diffuse model is sampled over a grid; as it would difficult to extrapolate, we simply used the galactic diffuse template as is.   Only photons with reconstructed energies between 60 MeV and 390 MeV are used in the fitting, and the photons in this energy range are fit in 14 log-spaced bins.
A fit extending to 390 MeV is necessary to improve the agreement between the fit and the model.  As in the case of LAT data, only front--converting photons are selected.  

\bibliography{refs1.bib}

\end{document}